\documentclass[manuscript]{aastex}
\usepackage{natbib}

\slugcomment{To appear in ApJL}

\shorttitle{Armchair PAHs and 12.7~$\mu$m band.}
\shortauthors{Candian et al.}

\begin{document}


\title{Polycyclic Aromatic Hydrocarbons with armchair edges and the 12.7 $\mu$m band.}


\author{A. Candian}
\affil{Leiden Observatory, Niels Bohrweg 2, 2333-CA, Leiden, The Netherlands}
\email{candian@strw.leidenuniv.nl}

\and
\author{P. J. Sarre}
\affil{School of Chemistry, The University of Nottingham, Nottingham, UK}

\and

\author{A. G. G. M. Tielens}
\affil{Leiden Observatory, Niels Bohrweg 2, 2333-CA, Leiden, The Netherlands}




\begin{abstract}
In this Letter, we report the results of density functional theory  calculations on medium-sized neutral Polycyclic Aromatic Hydrocarbon (PAH) molecules with armchair edges.
These PAH molecules possess strong  C-H stretching  and bending modes around 3~$\mu$m and in the fingerprint region (10-15~$\mu$m), and also strong ring deformation modes around 12.7~$\mu$m. Perusal of the entries in the NASA Ames PAHs Database  shows that ring deformation modes of PAHs are common, although generally weak. Therefore, we then propose that armchair PAHs with N$_C>$65 are responsible for the 12.7~$\mu$m aromatic infrared band in H{\sc ii} regions and discuss astrophysical implications in the context of the PAH life-cycle.

\end{abstract}


\keywords{astrochemistry - infrared: ISM - ISM: lines and bands - ISM: molecules - line: identification -  molecular data }



\section{Introduction}

Polycyclic Aromatic Hydrocarbon (PAH) molecules are commonly accepted to be responsible for a family of emission bands, the so-called Aromatic Infrared Bands (AIBs), which are seen in  a wealth of astronomical environments: H{\sc ii} regions, photodissociation regions (PDRs), planetary  and reflection nebulae, star forming regions, young stellar objects, galactic nuclei and (Ultra) Luminous Infrared Galaxies \citep{Hony01,Vand04,Tiel08}. However, despite much effort not a single PAH molecule has been identified unequivocally. Understanding the different components of the PAH population in terms of their molecular structure, charge and (de)hydrogenation would allow us to use them as a powerful tool to investigate the physical and chemical properties of astronomical objects \citep{Tiel08}.

The 11-15~$\mu$m region shows a host of features which have been attributed to C-H out-of-plane (OOP) bending modes \cite{Hony01}. The strongest of these bands peak at 11.2~$\mu$m and 12.7~$\mu$m and they come together with weaker ones at 11.0, 12.0, 13.5 and 14.2~$\mu$m \citep{Hony01}; all of them are perched upon a featureless and broad plateau \citep{Tiel08}.
The 12.7~$\mu$m band, due to the blending with atomic lines (Hu$\alpha$  at 12.37~$\mu$m, [NeII] at 12.8~$\mu$m) and molecular lines (H$_{2}~0-0$ S(3) at 12.3~$\mu$m)  was not studied in much detail before the advent of the Infrared Space Observatory \citep{iso}. In contrast to the red shaded asymmetric profiles of the 6.2 and 11.2~$\mu$m bands, the 12.7~$\mu$m band is characterised  by a slow blue rise and a red steep side \citep{Hony01}; its intensity with respect to the nearby 11.2~$\mu$m band varies by almost an order of magnitude according to object, reaching its relative maximum in H{\sc ii} regions \citep{Hony01}. 
Regardless of their charge, PAHs with duo or trio H show a band near 12.7~$\mu$m due to C-H OOP bending modes. As OOP modes are generally weak relative to the CC modes in cations \citep{Hudg94}, the 12.7~$\mu$m band is attributed to the OOP modes in neutral molecules containing duo or trio hydrogens \citep{Hony01, Baus08, Baus09, Ricc12}. Here we reexamine the assignment of the 12.7~$\mu$m feature based on the results of new quantum chemical calculations.

This Letter is organised as follows: Section 2 contains details of the specific PAH structures studied and the computational methods used; Section 3 discusses the results of the calculations, and in  Sections 4 and 5 the astrophysical implications and the conclusions are presented.

\section{Computational details}
There are several ways to classify PAHs, mostly based on the structure of their carbon skeleton and on the number of adjacent peripheral hydrogen atoms. If we consider PAHs as small sized graphene flakes,  they can be divided according to the shape of their edges. Thus we have PAHs with armchair edges like chrysene and PAHs with zigzag edges like tetracene (Figure 1(a)). PAHs with zigzag edges possess mostly solo (non-adjacent) hydrogen atoms, while PAHs with armchair edges possess mostly duo hydrogens which suffer from steric hindrance in the so called  bay regions (Figure (1a))

Density functional theory  has  proven to be a powerful tool to study the vibrational spectra of large molecules such as PAHs \citep{Lang96, Path05, Path07, Baus08, Baus09}.  In this study, the popular Becke three-parameter \citep{b3}, Lee-Yang-Parr \citep{lyp} B3LYP functional was used in conjunction with the 4-31G basis set \citep{Fris84} on Gaussian 03 \citep{g03} to optimize molecular geometries and to compute harmonic vibrational spectra. Langhoff (1996) showed that a scaling factor, 0.958, is needed to bring theoretical harmonic frequencies into agreement with experimental frequencies; this holds also for PAHs with armchair edges \cite{Lang96}. The spectra  are then  scaled and convolved with Lorentzian profiles of  30, 20 and 10 cm$^{-1}$ full-width half-maximum (FWHM) in the 3.1-3.4, 6-9, and 9-15~$\mu$m regions, respectively. 
Theoretical IR spectra are in absorption while astronomical IR spectra are in emission; commonly a redshift of 15~cm$^{-1}$ is invoked to account for anharmonicity in the emission process \citep{Baus08}. Given the lack of robust constraints on the value of the shift, in this study we chose to be conservative  and we do not apply a redshift to the theoretical spectra.
The software GABEDIT \citep{gabedit} was used to visualize the results of the calculations, and in particular to view the vibrational modes. The molecular structures studied are summarised in Figure 1(b): they were constructed to have increasing number of bay regions while retaining D$_{2h}$ symmetry.

\section{Results \& Discussion}
The theoretical IR spectra of the studied PAHs between 3 and 15~$\mu$m are shown in Fig.~2 (upper and middle rows). They all show strong bands around 3.25~$\mu$m  and in the fingerprint (11-15~$\mu$m) region caused by C-H stretches and bends, as is typical for neutral PAHs \citep{Lang96}. 
The pyrene-like PAHs (Fig.~2, upper row, first panel) possess several C-H stretching modes resulting in two strong peaks at 3.23 and 3.26~$\mu$m, in agreement with previous studies \citep{Baus09, Cand12}. The major contribution to the first peak is due to the symmetric stretching of C-H bonds involving duo hydrogens, while the second is the result of asymmetric stretching of C-H bonds involving  duo hydrogens and trio hydrogens. Increasing the size of the molecule results in more pronounced peaks, with the relative intensity  I$_{3.26}$/I$_{3.23}$ varying from 0.6 to 0.8 and a blue shift of the 3.23~$\mu$m peak. Perylene-like PAHs (Figure 2, middle row, first panel) also possess a double-peaked band in the 3~$\mu$m region that behaves similarly. 

In pyrene-like PAHs, the 6-9~$\mu$m region appears devoid of strong features (Figure 2, upper row, second panel). However, for perylene-like structures (Fig.~2, second row, middle panel) two moderately strong bands occur at $\approx$~6.41 and 7.21~$\mu$m, the former due to C-C stretching and the latter to a concerted in-plane bending motion of C-H bonds and stretching of C-C bonds. The positions of the bands appear independent of size, while the intensity increases with PAH size as can be expected for the larger number of bonds involved in the mode.

In the fingerprint region (10-15~$\mu$m) of pyrene-like PAHs (Figure 2, upper row, third panel) two peaks are noticeable around 12.0 and 12.7~$\mu$m. The peak at shorter wavelength is due to C-H OOP bending mode in duo Hs -- in agreement with accepted assignments \citep{ATB89}. Visualisation of the atomic displacements proves that the peak at 12.7~$\mu$m is the results of two equally strong, almost overlapping transitions due to a duo C-H OOP bending mode and a ring deformation mode (Table 1 and Figure 3).  This last vibrational mode is shown as a movie in Figure 3. It involves the rings belonging to the upper part  of the armchair structure; the central ``lone" rings  keep the structure flexible, thus also preserving the intensity of the band  for longer molecules.  The  intensity of the deformation mode increases steadily with number of carbon atoms and/or size as for the C-H OOP mode (Table 1).

In perylene-like molecules (Figure 2, middle row, third panel), the fingerprint region is more complex; the peak around 12.7~$\mu$m originates again from the C-C ring deformation mode and the two peaks at $\approx12.23~\mu$m and $\approx13.31~\mu$m are due to duo and trio C-H OOP bending modes, respectively. Weak modes  present at shorter wavelengths are due to a mix of C-H in plane and C-C bending modes. 

The infrared spectra of positively charged C$_{26}$H$_{14}$ and C$_{30}$H$_{16}$ were  computed to verify the effect of ionisation on the ring deformation modes around 12.7~$\mu$m. The band suffers a pronounced decrease in the intensity (Figure 2, lower row, third panel and Table 1), while its position shifts to shorter wavelengths; this is an effect which is generally true for all cationic PAHs compared to their neutral counterparts \citep{Hudg94}.  Vibrational modes in other regions of the spectrum behave as expected (Figure 2, lower row).

 We then inspected the NASA AMES PAH Theoretical Database \citep{Baus10, Boer14} to check whether the ring deformation mode around 12.7~$\mu$m was present in other neutral and charged PAHs.  We found that all PAHs in the database with armchair structure  possess a ring deformation mode in the range 12.3-12.8~$\mu$m. Also PAHs with a non-rigid central ring, such as dicoronylene (C$_{48}$H$_{20}$, uid$=$100), show a deformation mode around 12.55~$\mu$m.
The intensity of the ring deformation mode strongly depends on the flexibility of the carbon structure and on the degree of symmetry. It reaches a maximum in the molecules analysed in this Letter (Figure 1(b)), where the (series) of lone central rings facilitates the ring deformation mode without disrupting the symmetry; a lower symmetry results in decoupling of the vibrational modes. As the carbon structure becomes more rigid, e.g. in the case of C$_{40}$H$_{18}$ (uid$=$555), the intensity of the deformation mode  around 12.59~$\mu$m rapidly decreases and it becomes very weak in pericondensed PAHs with armchair edges like  C$_{36}$H$_{16}$ (uid$=$128). 

\section{Astrophysical implications}

While the 12.7~$\mu$m band is seen in a wide range of astronomical environments, it appears to be particurlarly prominent in the IR spectra of H~{\sc ii} regions, where its strength can be comparable to that of the 11.2~$\mu$m band \citep{Hony01}. Likewise, the 12.7~$\mu$m band is very strong in the surface layers of PDRs where the 6.2 and the 7.7~$\mu$m bands are also (relatively) strong.
Our calculations show that neutral PAHs with armchair edges and/or central flexible ring(s) possess a ring deformation mode in the 12.3-12.7~$\mu$m range, where the intensity depends on the flexibility of the molecule and its degree of symmetry.
These PAH molecules can thus make an important contribution to the 12.7~$\mu$m band strength, especially in H {\sc ii} regions.
However, to reproduce the relatively low 11.2/12.7~$\mu$m observational intensity ratio, these molecules should not possess solo (non-adjacent) hydrogen atoms. Indeed, the intrinsic intensity of the solo OOP bending mode is higher than the intensity of a ring deformation mode \citep{CandT}.\footnote{Visualisation of the vibrational modes of dicoronylene with the NASA AMES PAH Database online tool shows a strong (126.1 km/mol) solo OOP bending mode at 11.19~$\mu$m, compared to the deformation mode at 12.55~$\mu$m (26.3 km/mol).}  Thus, the interstellar 11.2 and 12.7~$\mu$m bands must be carried by separate populations of PAHs. In contrast, within this framework, the 12.0 and 12.7 um bands would result from these armchair PAHs. In Figure 4 we compare the 12.0/12.7~$\mu$m ratio as function of size (N$_c$) in pyrene-like molecules (squares) with the values measured in H{\sc ii} regions (dotted lines) by Hony et al. (2001). This comparison reveals that PAHs in excess of 65 C atoms are needed  to reproduce the observed low ratio. This is in agreement with the size estimate for interstellar PAHs derived from the observed 3.3/11.2~$\mu$m ratio \citep{Tiel08, Ricc12}.

The observed good correlation between the 12.7~$\mu$m band and the 6.2~$\mu$m band has always presented an enigma for the PAH model as the former is invariably attributed to neutral PAHs while the latter is carried by ionized PAHs. Generally, this correlation is taken to imply that the conditions that are leading to ionization of interstellar PAHs also promote a high abundance of PAHs with corners (\textit{i.e.}, PAHs with duo and trio H atoms; \cite{Hony01, Ricc12}). This may then imply that PAHs are rapidly broken down in the ionization zones and the fragmentation process produces small PAHs that carry the 12.7~$\mu$m band. However, as the lifetime of small PAHs is expected to be much shorter than for large PAHs, it is unlikely that this interpretation can withstand a quantitative analysis. Moreover, the main destruction channel for interstellar PAHs is likely initiated by complete dehydrogenation and the formation of pure carbon graphene flakes and/or cages \citep{Eker98, Jobl04, Zhen14, Bern12}. Therefore, breakdown of interstellar PAHs will not favour the formation of the 12.7~$\mu$m carriers. The identification of the 12.7~$\mu$m band with armchair PAHs proposed here shines new light on this problem. In particular, PAHs with armchair edges are known to be much more stable than PAHs with zigzag edges \citep{Poat07}. Thus, in the surface layers of PDRs, UV photons lead to rapid ionization while also reducing the interstellar PAH family to its most stable form, the armchair PAHs. Theoretical studies have shown that carbon loss (C$_2$H$_n$) from large compact PAHs leads to the formation of armchair  structures \citep{Baus14}.   While this class of PAH molecules can describe the qualitative behaviour of the observed 12.7/6.2 ~$\mu$m bands, experimental studies on the photochemical evolution of large PAHs are  needed to establish their survival rate and to quantify the abundance of elongated armchair PAH molecules in the ISM.

 
Candian et al (2012) proposed that the 3.3~$\mu$m C-H band of class A -- as observed in H {\sc ii} regions  --  can be explained in terms of a two-component population of PAHs: compact molecules and molecules with bay regions. The PAHs investigated here possess bay regions and show two strong peaks in the C-H stretch region (Figure 3). They can thus contribute to the bay  component responsible for the short-wavelength part of the 3.3~$\mu$m band. An interesting way to explore this hypothesis would be follow the spatial variation of the two components of the 3.3~$\mu$m band and of the 12.7~$\mu$m band in a sample of PDRs.

\section{Conclusions}

In this Letter we studied the vibrational spectra of PAHs with armchair edges by means of  Density Functional Theory. The results presented here demonstrate that coupling between the C-H OOP mode and the C-C ring deformation mode in this sub-class of species lead to a strong 12.7~$\mu$m band. Hence we propose that armchair PAHs with N$_c> 65$ are the main carriers of the interstellar 12.7~$\mu$m band, previously attributed to duo and trio C-H out-of-plane bending modes. 
It is inferred that PAHs with armchair edges are favoured in regions exposed to strong UV processing.
 In the future, the full coverage of the AIB spectrum attained by the \textit{James Webb Space Telescope} will allow us to study the spatial variation of single AIB features with unprecedented sensitivity, helping us to further understand the contribution of PAHs with armchair edges to the astronomical PAH population.

\acknowledgments
AC acknowledges STFC and The University of Nottingham for scholarships and Cameron Mackie for helping with the animation. Studies of interstellar PAHs at Leiden Observatory are supported through advanced European Research Council Grant 246976 and a Spinoza award. The calculations were performed at the High Performance Computer (HPC) facility at The University of Nottingham (UK) and at the SARA supercomputer center in Amsterdam, The Netherlands (project MP-270-13).

\clearpage

\begin{figure}
\includegraphics[scale=0.68]{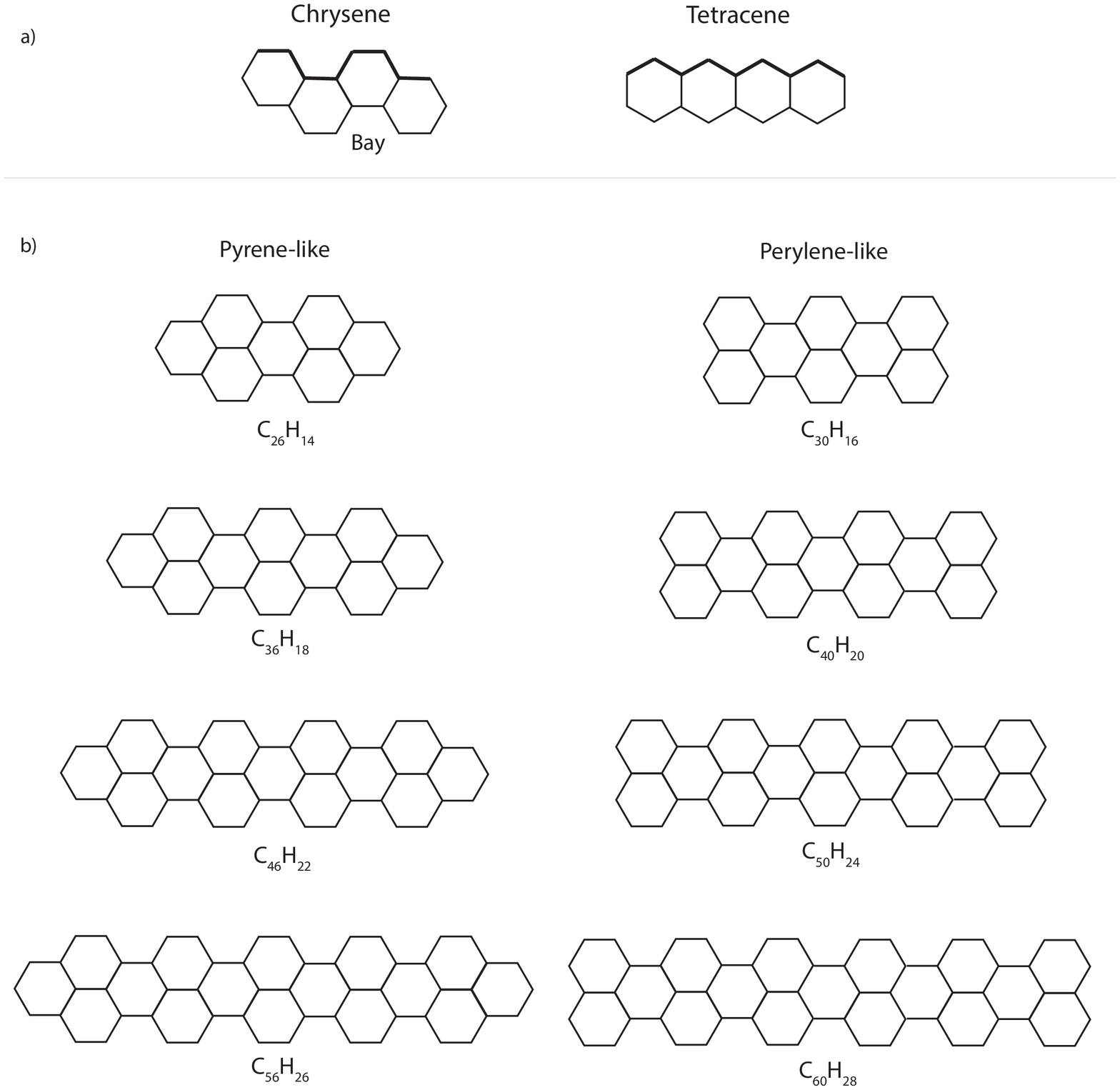}
\caption{(a) Examples of isomeric PAH molecules (C$_{18}$H$_{12}$) with different edges. Chrysene (left) possesses armchair edges, while tetracene (right) has zigzag edges. Note that PAHs with armchair edges have bay regions. (b) Molecular carbon structures of studied PAHs with armchair edges in order of size. These moelcules can be described as pyrene-like  (ending with a group of three adjacent hydrogens, \textit{e.g.} C$_{26}$H$_{14}$) and perylene-like (ending with two groups of three adjacent hydrogens, \textit{e.g.} C$_{30}$H$_{16}$). All of them belong to the D$_{2h}$ point group.}
\end{figure}

\clearpage

\begin{figure}
\includegraphics[scale=.80]{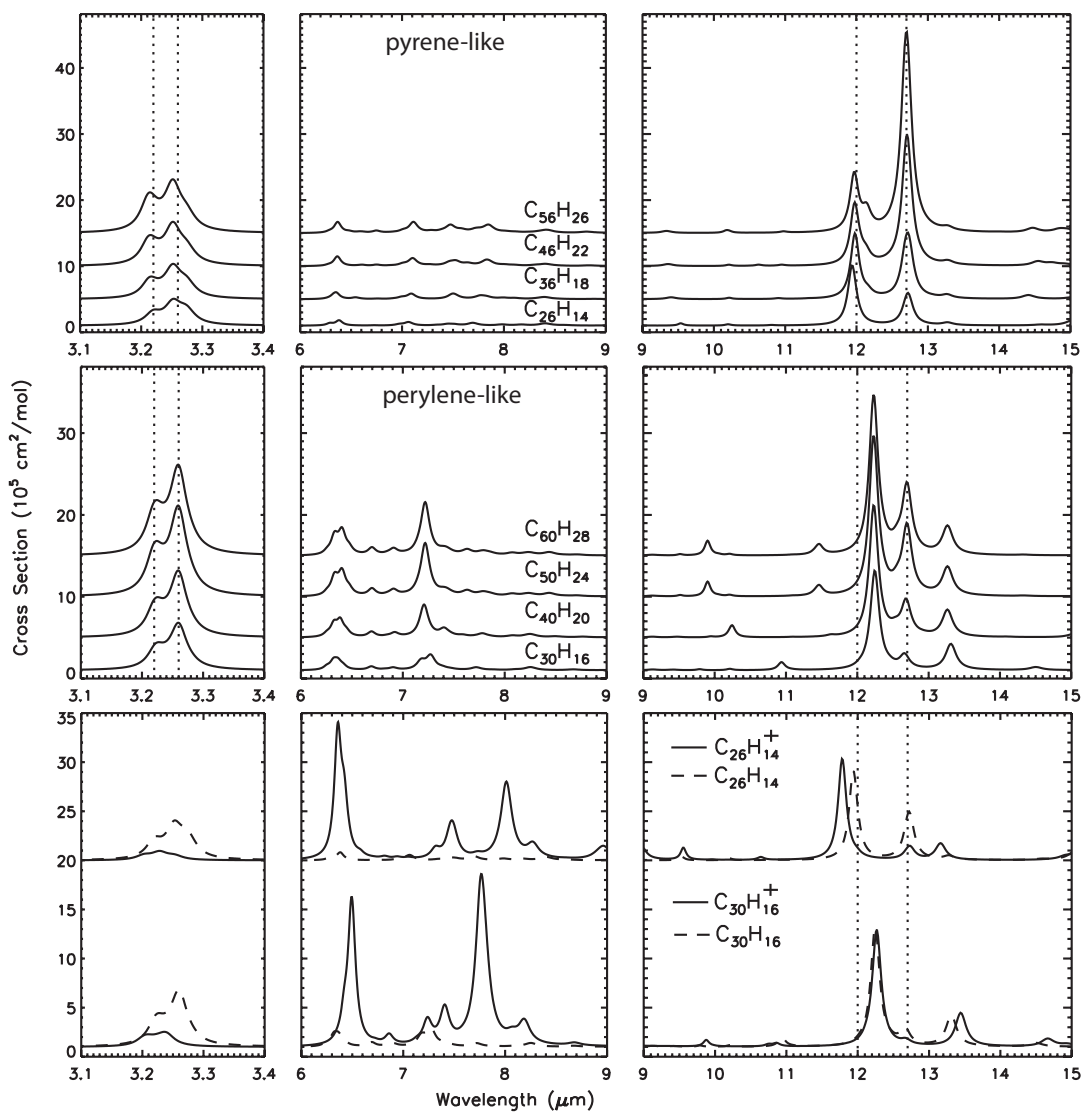}
\caption{ Theoretical absorption IR spectra of the molecular structures of Figure 1(b). A Lorentzian broadening and a scaling factor are applied (see text for details). For ease of comparison, dotted lines are drawn at 3.22, 3.26, 12.0 and 12.7~$\mu$m. The upper row shows the spectra of pyrene-like structures, the middle row the spectra of perylene-like structures and the lower row a comparison  between the IR spectrum of the first member of each series (dashed line), and the spectrum of its positively charged counterpart (solid line).}
\end{figure}

\clearpage

\begin{figure}
\includegraphics[scale=0.7]{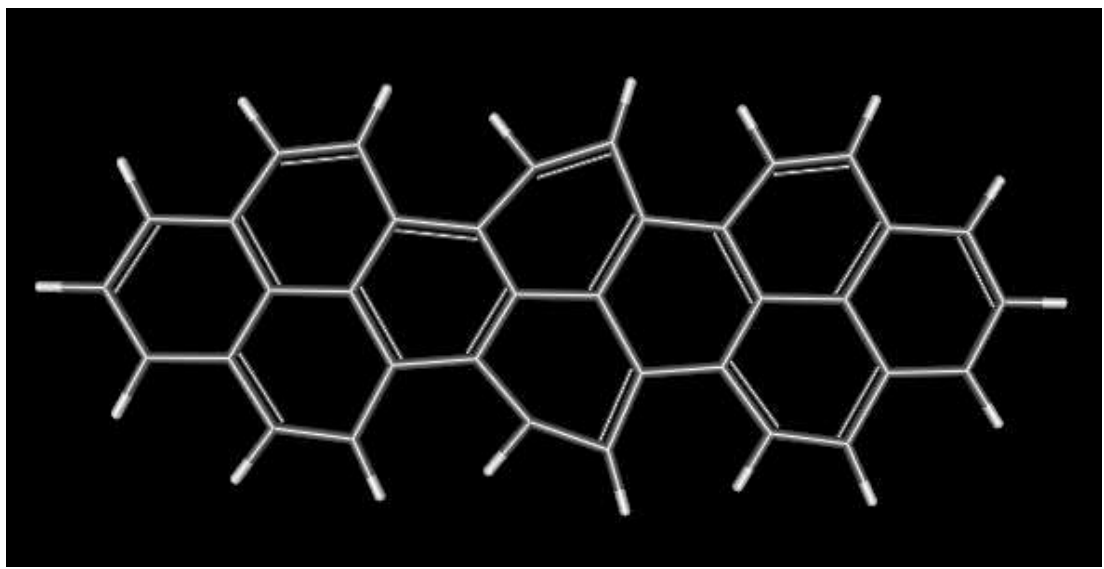}
\caption{Visualisation of the vibrational mode occurring around 12.7~$\mu$m in C$_{26}$H$_{14}$. The carbon skeleton is shown in grey, the hydrogen atoms in white, and the arrows represent the motion direction and intensity. The ring deformation mode, belonging to the B$_{1u}$ symmetry, involves rings forming the upper part of the armchair edge. 
(An animation of this figure is available in the online journal.}
\end{figure}

\clearpage

\begin{figure}
\epsscale{0.8}
\plotone{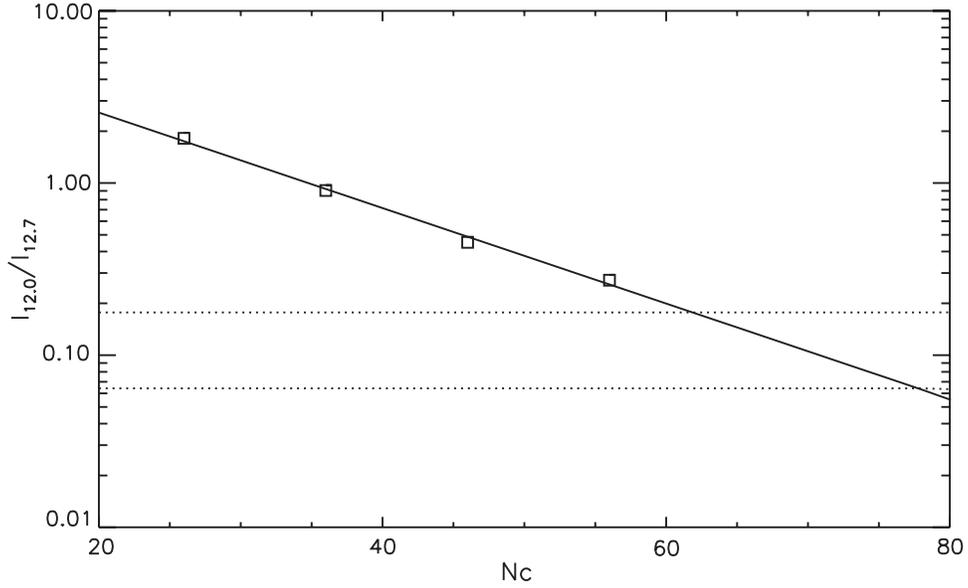}
\caption{Ratio of the theoretical intrinsic intensity of the 12.0~$\mu$m band (C-H OOP bends) and the 12.7~$\mu$m band (duo C-H OOP bends+ C-C deformation mode)  as function of the number of carbon atoms for the pyrene-like series. The solid line shows the result of a linear fit. Calculation of the intensity through an emission model would only slightly increase the ratio given the proximity of the 12.0 and 12.7~$\mu$m bands. The horizontal dotted lines indicates the observed astronomical range in H{\sc ii} regions as measured by Hony et al (2001).}
\end{figure}

\clearpage

\clearpage
\begin{table}
\begin{center}
\caption{Vibrational modes in the 12-14~$\mu$m Region for the molecules studied}
\begin{tabular}{cccc|cccc}
\tableline\tableline
 &Pyrene-like                               &      &          &  & Perylene-like                             &      & \\

Molecule & $\mu$m (cm$^{-1}$) & Int & Type & Molecule & $\mu$m (cm$^{-1}$) &  Int & Type\\
\tableline
\tableline
C$_{26}$ H$_{14}$ &   11.93 (837.6) & 140.9 & C-H & C$_{30}$ H$_{16}$ &  12.25 (816.6) & 190.3 & C-H\\
                              &  12.71 (787.0) &   23.1 & C-C &                               &   12.66 (789.8) &  25.3  & C-C \\
                              &   12.72 (785.8) &   54.4 & C-H &                              &   13.31 (751.2) &  49.4  & C-H \\
C$_{36}$ H$_{18}$ &  12.00 (834.9) & 153.5 & C-H & C$_{40}$ H$_{20}$ &   12.23 (817.6) & 246.2 & C-H \\
                              &   12.67 (788.1) &   73.0 & C-C &                              &   12.68 (788.4) &  64.8  & C-C \\
                              &   12.74 (785.2) &   96.7 & C-H &                              &   13.26 (754.0) &  50.7  & C-H \\
C$_{46}$ H$_{22}$ &   11.97 (835.2) & 141.3 & C-H &C$_{50}$ H$_{24}$ &   12.23 (817.7) & 304.5 & C-H\\
                              &   12.70 (787.2) & 148.0 & C-H &                              &   12.70 (787.6) & 129.2 & C-C\\
                              &  12.71 (786.7) & 164.6 & C-C &                               &   13.26 (753.9) &   52.9 & C-H\\
C$_{56}$ H$_{26}$ &   11.97 (835.6) & 133.8 & C-H & C$_{60}$ H$_{26}$ &  12.23 (817.5) & 363.0 & C-H\\
                              &   12.68 (788.4) & 200.4 & C-H &                               &  12.71 (787.1) & 225.9 & C-C\\
                              &  12.71 (786.7) & 290.5 & C-C  &                               &   13.28 (753.2) &  50.7 & C-H\\
\tableline
C$_{26}$ H$_{14}^+$  & 11.78 (848.7) & 162.2 & C-H  & C$_{30}$ H$_{16}^+$ & 12.27 (815.2) & 186.5  & C-H\\
                                   & 12.68 (788.9) &   1.7 & C-C &                                      &  12.68 (788.9) & 6.86    & C-C\\
                                   & 12.73 (785.5) & 20.1 & C-H &                                     &  13.44 (743.8)  & 53.36  & C-H\\ 
\end{tabular}
\tablenotetext{}{Frequencies are scaled. Infrared intensities are in km mol$^{-1}$.}
\end{center}
\end{table}

\clearpage

\end{document}